# A Performance Analysis of LAR Protocol for Vehicular Ad Hoc Networks in City Scenarios

Sanjoy Das[1], and D.K Lobiyal[2]
[1,2]School of Computer and Systems Sciences
Jawaharlal Nehru University, New Delhi, India
Email: [1]sdas.jnu@gmail.com and [2]lobiyal@gmail.com

*Abstract*—In this paper, performance analysis of Location Aided Routing (LAR) protocol in different city scenarios has been done. The mobility model considered is Manhattan model. This mobility model used to emulate the movement pattern of nodes i.e., vehicles on streets defined by maps. Our objective is to provide a qualitative analysis of the LAR protocol in different city scenarios in Vehicular Ad hoc Networks. We have considered three different city scenarios for the analysis of the protocol. The simulation work has been conducted using the Glomosim 2.03 simulator. The results show that LAR1 protocol achieves maximum packet delivery ratio is 99.68 % and maximum average end-to-end delay is 7.319969 ms when the network is sparsely populated. Further, for densely populated network maximum achieved packet delivery ratio is 87.58% and average end-to-end delay is 0.017684 ms.

*Index Terms*— LAR, Vehicular Ad Hoc Network, Mobility Model, Manhattan model, Packet Delivery Ratio, End-to - End Delay

## I. INTRODUCTION

VANET is a special class of Mobile Ad hoc Network (MANET), where every node is a vehicle moving on the road. In this network a node behaves like a router to relay a message from one node to another. In VANET mobility of vehicles; structure of the geographical areas since node movement depends on it, timely delivery of messages, privacy are very important characteristics. VANET uses two types of communication methods- One from vehicle to vehicle (V2V) and the other is vehicle to fixed road side equipment (V2R). In both the methods vehicles can communicate to other vehicles or road side unit either directly or through multiple hops. This totally depends on the position of the vehicles [1]. Further, the road side units (RSU) can also communicate with other RSU via single or multi hop fashion. The RSU supports numerous applications like road safety, message delivery; maintain connectivity by sending, receiving or forwarding data in the network. The main focus of the VANET is to provide real-time and safety applications for drivers and passengers. By delivering message on time can minimize road accidents and save total journey time. The RSU can improve traffic management system by providing drivers and passengers with vital information i.e., collision warnings, road sign alarms, blind turn warning, etc. There are various services currently support by VANET are internet connections facility, electronic toll collection, and a variety of multimedia services. It is desirable that protocols should maintain the low end-to-end delay and, high delivery ratio, low overheads and minimum numbers of hops.

## II. RELATED WORKS

Extensive works have been carried out by researchers, academicians and industries for successfully routing of messages in VANET. There are several research projects [2], [3], [4], [5] on VANET being carried out by researchers. Some of them are [CarTalk, FleetNet–Internet on the Road, NoW (Network on Wheel)] with emphasis on deployment in the real world. The main focus of all these projects is to provide safety, and timely dissemination of message from one location to another location. Some of the message delivery protocols proposed for VANET attempt to deliver a message to a geographic region rather than to a node. These protocols are called geocast routing. LAR [6], LBM [7], and GeoTORA [8] is modified TORA, GRID protocol is modified to GeoGRID [9], DREAM [10], GRUV [11], are few geocasting protocols. In Ref. [12] authors have only considered the energy consumption parameter for performance analysis of LAR1 protocol with DSR and AODV in highly dense ad hoc networks. The results reported show that LAR1 perform better than DSR and AODV protocol in highly dense network. But in low density DSR performs better than others in term of energy consumption. In [17] authors show a wide analysis of their proposed protocol Geographic Source Routing (GSR) with DSR, AODV for VANET in city scenarios. They have done simulation analysis of these protocols on realistic vehicular traffic for a particular city. The real city map is considered and converted to graph for the analysis. Their result shows that GSR performs better than DSR and AODV in terms of end-to-end delivery and latency. In [13] and [18] the authors proposed different modified LAR algorithms. They have modified the request zone. Through simulation, the authors have established that their proposed algorithms reduces route request overhead as compared to original LAR. The performance analysis shows that their method outperforms original LAR especially, in a dense and highly dynamic ad hoc network. In [16] the authors have proposed a greedy version of LAR protocol known as GLAR (Greedy Location-Aided Routing Protocol). This scheme improved the performance of LAR. In GLAR method, to find a route between source and destination, a baseline is drawn between them. The route request packets are broadcast within the request zone. The neighboring node which has shortest distance towards baseline is selected as next broadcasting node. The authors considered various network performance parameters to compare LAR with GLAR. Their results revealed that GLAR reduces the number of route discovery packets and increases the average network route lifetime. Most of these protocols use random waypoint mobility model for performance analysis. The protocols proposed in [13], [16], [17], and [18] did not consider structured city scenarios for the performance analysis of LAR1 protocol in VANET.

### III. OVERVIEW OF LAR PROTOCOL

Y.B. Ko et.al in [6] proposed two different location aided schemes for transmitting a message from source to destination known as LAR scheme 1 and LAR scheme 2. Both the schemes used the location information of source and destination nodes to reduce the routing overhead. It assumes that the local geographic information is obtained using the global positioning system. In LAR scheme 1, an expected zone is computed for the possible position of the destination node. It is a circle around the destination that contains the estimated location of the destination node. The Request Zone is a rectangle with source node S in one corner (Xs, Ys), and the Expected Zone containing destination D in the other opposite corner $(X_d, Y_d)$. In this protocol, only those neighbours of source node that are present within the request zone forwards the route request packet further. The source node S know the location of destination node $D(X_d, Y_d)$ at time $t_0$ and average speed v with which D is moving. Every time node S initiates a new route discovery process, it the circular expected zone at time $t_1$ with the radius R = $v(t_1 - t_0)$ and center at location $(X_d, Y_d)$. In Fig.1, I and J are neighbours of Source node S. But, only node *I* forwards the packets received from S to its neighbours, since *I* is within the request zone. The node J discards the message received from S since J is outside the request zone.

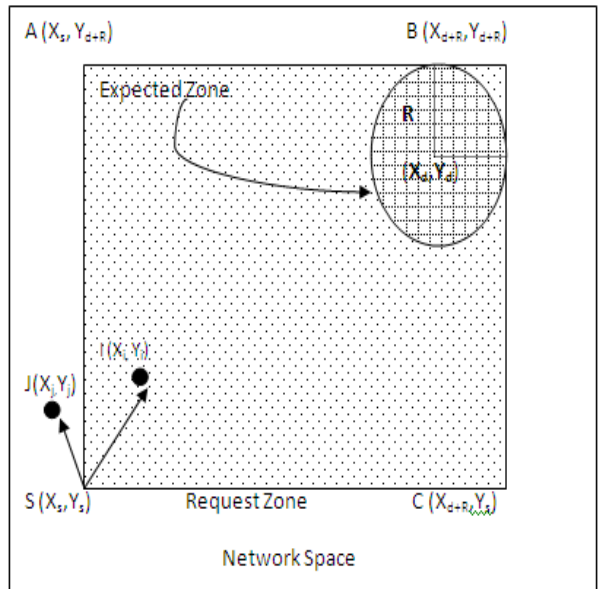

Figure 1. LAR1 Routing protocol [6]

### IV. DESCRIPTION OF MANHATTAN GRID MOBILITY MODEL

The Manhattan mobility model [15] imitates the movement patterns of mobile nodes on streets defined by maps shown in Fig.2. It is useful in modeling movement of nodes in an urban area or city scenarios. This model follows grid road topology. The map is divided into number of horizontal and vertical streets. Each street is divided into two lanes in each direction. In vertical Street nodes move north and south direction and horizontal streets nodes move in east and west direction. The horizontal and vertical street creates several intersection points. The nodes can take turn from intersection point in left, right or go straight direction. The model works based on a probabilistic approach for selection of nodes movements. The probability of moving in the same direction is 0.5 and the probability of turning left and right is 0.25.

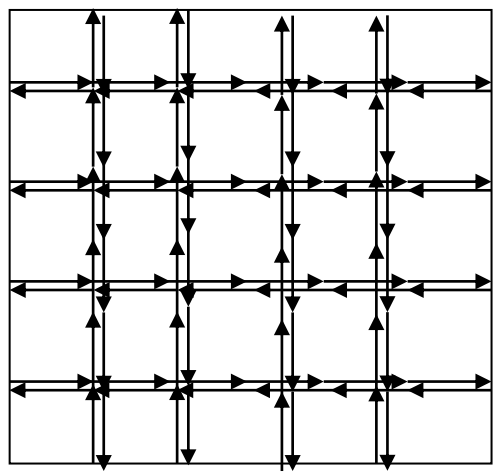

Figure 2. Manhattan model of 5x5 grid topology. Vehicular nodes move according to direction shown.

## V. SIMULATION ENVIRONMENTS AND RESULT ANALYSIS

The simulation has been carried out to evaluate the performances of LAR scheme 1 protocol in different city scenario for VANETs by using the network simulator Glomosim-2.03[14]. It is one of the widely used for research in mobile ad hoc networks and freely available simulator. The movements of nodes and scenarios are generated using mobility generator tool bonnmotion-1.5 [15].The table 1 shows different simulation parameters and table 2 shows the different parameters values considered for simulation. We considered three different mobility scenarios and parameter values are accordingly considered as shown in Table 3.Results for all three scenarios have been presented in table 4.In the results we have computed the average end to end delay and packet delivery ratio for all three scenarios. All the results presented are obtained average of 10 different simulation runs with different seed values.

TABLE I.
SIMULATION PARAMETERS

| Parameter | Specifications |
|---|---|
| MAC Protocol | IEEE 802.11 DCF |
| Speed | Uniform |
| Data Type | Constant Bit Rate (CBR) |
| Radio Propagation Model | Two-ray ground reflection model |
| Channel type | Wireless channel |
| Antenna model | Omni-directional |
| Routing Protocol | LAR1 |
| Mobility Model | Manhattan Grid Mobility Model |

TABLE II.
VALUES OF SIMULATION PARAMETERS

| Parameter | Values |
|---|---|
| Simulation Time | 1000s |
| Simulation Area | 1000x1000 |
| No of Nodes( Vehicles) | 25,50,75,100,125,150 |
| Data Packet Sizes | 512bytes |
| Transmission Range | 250 m |
| Mean speed | 10 |
| No. of blocks along x-axis | 10 |
| No. of blocks along y-axis | 15 |

TABLE III.
VALUES OF DIFFERENT SCENARIOS

| Scenario | Speed change probability | Min. speed (m/s) | Turn probability |
|---|---|---|---|
| 1 | 0.25 | 10 | 0.25 |
| 2 | 0.5 | 20 | 0.5 |
| 3 | 0.75 | 30 | 0.75 |

### A. Packet Delivery Ratio

Packet delivery ratio is a very important factor to measure the performance of routing protocol in any network. The performance of the protocol depends on various parameters chosen for the simulation. The major parameters are packet size, no of nodes, transmission range and the structure of the network. The packet delivery ratio can be obtained from the total number of data packets arrived at destinations divided by the total data packets sent from sources.

Packet Delivery Ratio = 
$$\frac{\Sigma(Total packets\ received\ by\ all\ destination\ node)}{\Sigma\ (TotalPacketssendbyallsourcenode)}$$

### B. Average End to End Delay

End-to-end delay is the time taken by a packet to route through the network from a source to its destination. The average end-to-end delay can be obtained computing the mean of end-to-end delay of all successfully delivered messages. Therefore, end–to-end delay partially depends on the packet delivery ratio. As the distance between source and destination increases, the probability of packet drop increases. The average end-to-end delay includes all possible delays in the network i.e. buffering route discovery latency, retransmission delays at the MAC, and propagation and transmission delay.

Fig.3 shows the packet delivery ratio is nearly 99.68 % for a network size of 25 nodes and 84.38% for a network 150 nodes. The PDR is slowly decreased as the numbers of nodes increases in the network. One abnormal behavior noticed that when number of nodes is 125 the PDR value increases. The PDR value increases due to fewer collisions in the network.

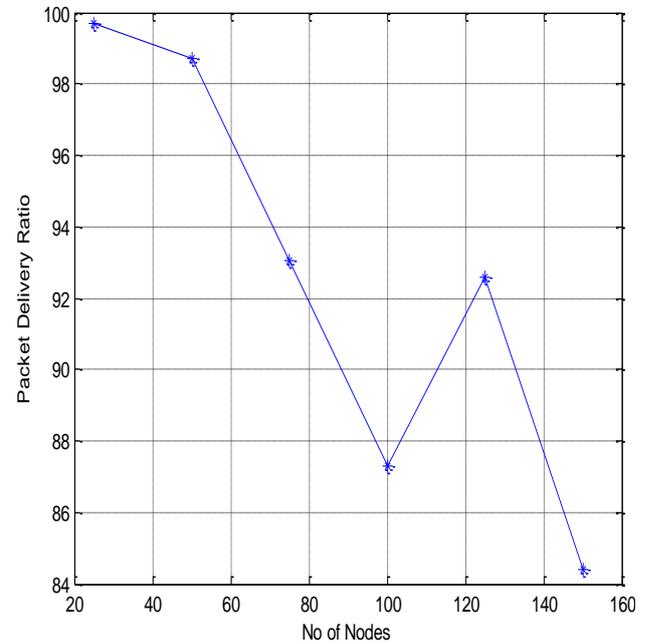

Figure 3. Packet Delivery Ratio for scenario 1

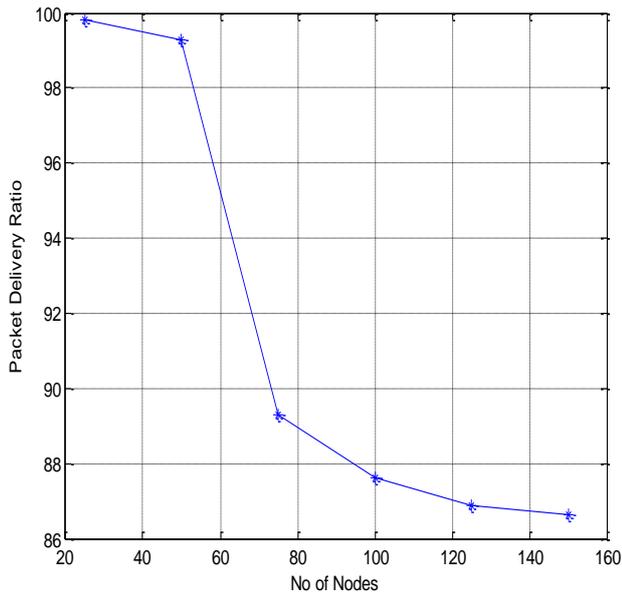

Figure 4. Packet Delivery Ratio for Scenario 2

Fig.4 shows the packet delivery ratio is nearly 99.28 % for network size of 25 nodes and 86.64% for 150 nodes. The PDR is suddenly decreased from 99.28% to 89.28% when number of nodes increased from 50 to 75. As the number of nodes increases from 75 to 150, the PDR decreases slower. .

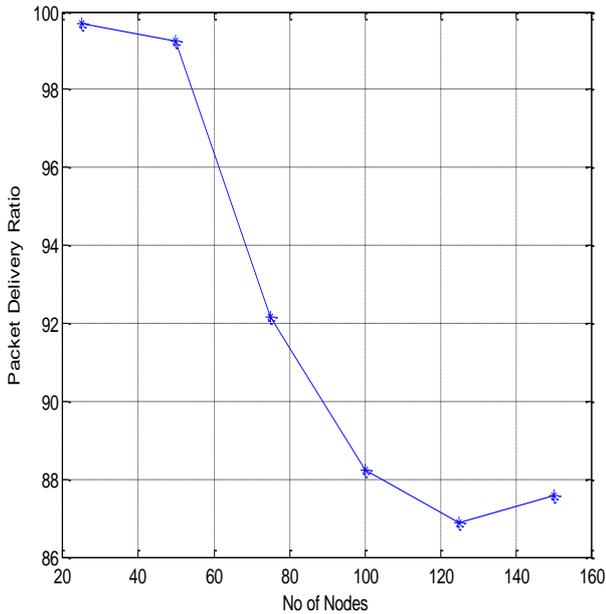

Figure 5. Packet Delivery Ratio for Scenario 3

Fig.5 shows that packet delivery ratio is nearly 99.24 % for the network size of 25 nodes and 87.57 % for 150 nodes. The PDR is suddenly decreased from 99.24% to 92.15% when number of nodes increased from 50 to 75. As the number of nodes increases from 75 to 125, the PDR decreases slowly. For number of nodes 150 there is slight increase in the PDR.

Fig.6 shows that average end- to- end delay is nearly 5.938166 for the network size of 25 nodes and 0.017684 for 150 nodes. The average end-to- end delay is suddenly decreased from 5.938166 to 1.7990836 as the number of nodes increased from 25 to 50. When the number of nodes increases from 75 to 150, the average end- to- end delay decreases slowly.

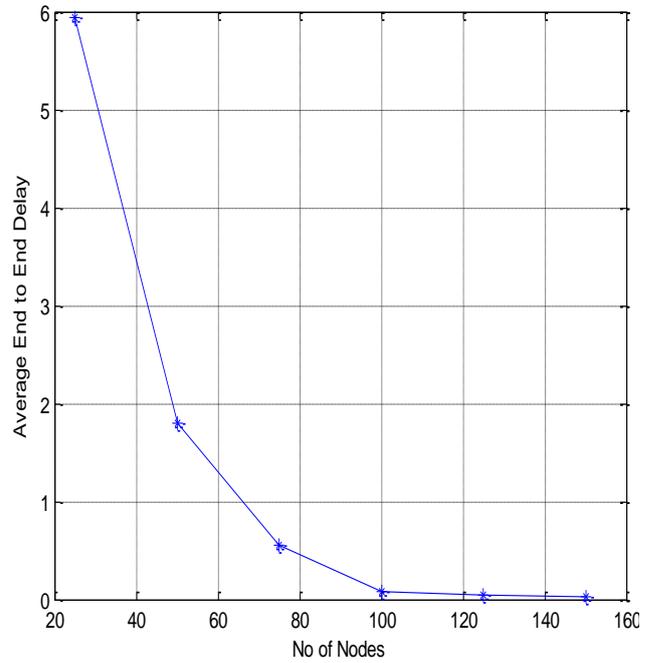

Figure 6. Average End to End Delay for Scenario 1

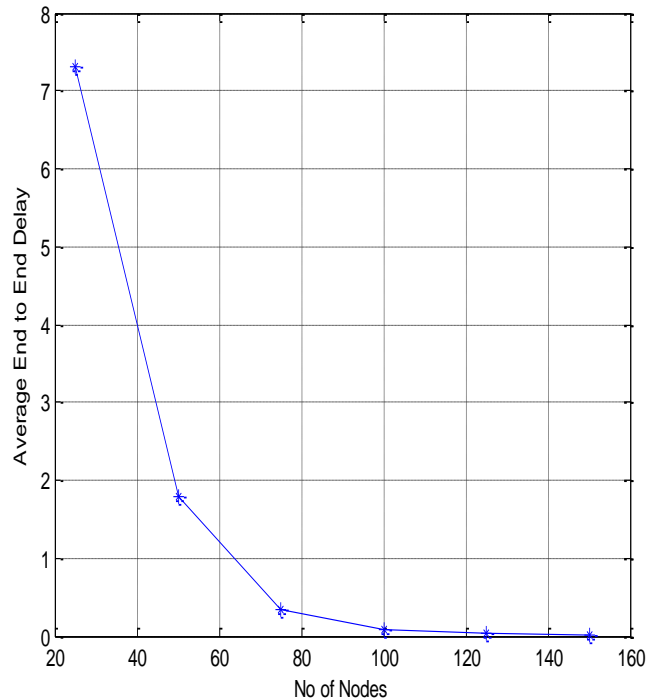

Figure 7. Average End to End Delay for Scenario 2

Fig.7 shows that average end–to-end delay is nearly 7.319969 for the network size of 25 nodes and 0.016942 for 150 nodes. In the scenario 1, we have observed that the average end- to- end delay for the network size of 25 nodes is 5.938166. But in this scenario, the value is increased. The average end–to-end delay is suddenly decreased from 7.319969 to 1.785661 as number of nodes increased from 25 to 50. For the network size of 50 nodes, the value of end-to-end delay is lower as compared to scenario 1. As the number of nodes increases from 75 to 150, the average end-to-end delay decreases.

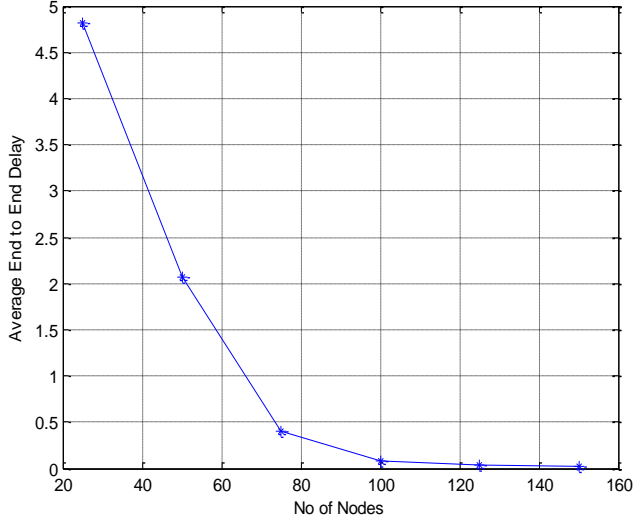

Figure 8. Average End to End Delay for Scenario

Fig 8 shows that average end-to-end delay is nearly 4.810926 for the network size of 25 nodes and 0.016494 for 150 nodes. The average end–to-end delay is suddenly decreased from 4.810926 to 2.074692 as the number of nodes increases from 25 to 50. As the number of nodes increases from 75 to 150, the average end-to-end delay decreases.

TABLE IV.
PDR AND AVERAGE END-TO- END DELAY OF ALL SCENARIOS

| Metric | Scenario | No of nodes | | | | | |
|---|---|---|---|---|---|---|---|
| | | 25 | 50 | 75 | 100 | 125 | 150 |
| Average End to End Delay (m/s) | 1 | 5.938166 | 1.7990836 | 0.551479 | 0.084834 | 0.0385 | 0.017684 |
| | 2 | 7.319969 | 1.785661 | 0.331006 | 0.082103 | 0.037262 | 0.016942 |
| | 3 | 4.810926 | 2.074692 | 0.404131 | 0.0837 | 0.039286 | 0.016494 |
| Packet Delivery Ratio (%) | 1 | 99.68 | 98.7 | 93.04 | 87.32 | 92.584 | 84.38 |
| | 2 | 99.8 | 99.28 | 89.28 | 87.63 | 86.872 | 86.64 |
| | 3 | 99.68 | 99.24 | 92.14667 | 88.25 | 89.064 | 87.58667 |

## VI. CONCLUSIONS

In this paper we have analyzed the performance of LAR1 protocol for vehicular ad hoc networks in city scenario with Manhattan Mobility model. The performance of the protocol is thoroughly studied with varying node density, node speed and various parameters of mobility model. We have calculated packet delivery ratio and end-to-end delay for different scenarios. From the result analysis it is clearly evident that when the network is sparsely populated the successful delivery of message is nearly 99%. The end–to-end delay is high in sparsely populated network but in densely populated network end–to-end delay is low. This protocol may be used in sparse traffic where delay does not affect the performance of the network. Further it may be used in densely traffic condition where PDR is not important.